\documentclass[prb,twocolumn,showpacs]{revtex4}

\usepackage{graphicx}
\usepackage{amsmath}
\usepackage{amssymb}
\usepackage{epsf,latexsym}
\usepackage{times}

\newcommand{\be}{\begin{equation}}
\newcommand{\ee}{\end{equation}}
\newcommand{\ba}{\begin{eqnarray}}
\newcommand{\ea}{\end{eqnarray}}

\newcommand{\putfig}[2]{$$\leavevmode\hbox{\epsfxsize=#2 cm
   \epsffile{#1.eps}}$$}

\newcommand{\Fig}[1]{Fig.~\ref{#1}}
\newcommand{\Eq}[1]{Eq.~(\ref{#1})}

\newcommand{\Refon}[1]{Ref.~\onlinecite{#1}}

\begin{document}


\title{Quantum buses and quantum computer architecture based on quantum dots}

\author{Irene D'Amico}
\email{ida500@york.ac.uk}

\affiliation{Department of Physics, University of York, York YO10,United Kingdom}

\date{\today }

\begin{abstract}
We propose a  quantum computer architecture based on quantum dots  both for short distance  and for long distance communication/computation. Our scheme exploits the natural characteristics of self-assembled quantum dots and it is  scalable.
It is centered on the idea of  a
quantum bus based on semiconductor self-assembled quantum dots. This allows for transmission of qubits between  the different quantum registers, and could be integrated in  most of the present proposal for
  semiconductor quantum dot-based quantum computation. Our proposal exploits the peculiar properties of {\it relatively short} spin-chains, and advantages and disadvantages of two possible 
implementations, both based on spin-chain global dynamics, are discussed in details.  
A clear advantage of the scheme is to avoid the use of  microcavities for long distance communication  
between different elements of the quantum computer. In this respect our scheme is comparatively faster than hybrid quantum dot-microcavity schemes.
\end{abstract}




\maketitle 

\section{Introduction}
In the past years, self-assembled semiconductor quantum dots\cite{QD} have been considered one of the most promising hardware for envisaging quantum information/computation (QIC) schemes (see e.g. \Refon{PRBBiolatti,PRBSergio,Briggs,Mang1,Mang2,Paulibl,Imamoglu}). The key feature of this hardware is the possibility of optically drive quantum calculations by multicolor trains of laser pulses, overcoming in this way the relatively fast decoherence times typical of solid state systems. Another important characteristic is  the fact that, since self-assembled quantum dots (QDs) tend to be naturally produced in large arrays, such a hardware could in principle solve  the problem of scalability, which affects other types of implementations.

Many QD-based QIC schemes  consider as quantum arrays ensembles of vertically stacked quantum dots\cite{PRBBiolatti,PRBSergio,Briggs,Mang2,Paulibl}, which spontaneously form in  multilayer QD structures\cite{QD}.
However following the initial enthusiasm, it has become clear that this choice solves only partially the scalability problem. Experiments show in fact that the maximum length of such vertical arrays
is of the order of ten quantum dots only. Clearly this rises the problem 
 of how to perform complex computations or equivalently, how to communicate between different arrays.
A possible solution for communicating between far away quantum dots/arrays is to embed the system in a microcavity and then to use the cavity mode to transmit qubits of information between well separated quantum dots. This exploits the potential conversion of trapped excitons into flying qubits and vice-versa\cite{Imamoglu,Mang1}. Among the problematic issues related to this scheme are (i) to experimentally embed self-assembled arrays of quantum dot in a high Q microcavity, (ii)  to carefully position the dots away from the cavity mode nodes in order to preserve QD-cavity coupling, (iii) to engineer the structure such that it is still possible to address the QDs by laser fields and possibly couple each QD with a static external electric fields, as required, e.g., in \Refon{PRBBiolatti,PRBSergio,Paulibl}. Experimentally this is very demanding.

To lift these complex requirements, we propose to  communicate between far away vertical arrays of QDs by using quantum buses made by {\it in-plane} chains of quantum dots. To implement this, our proposal exploits the structural differences and peculiar  characteristics of in-plane versus vertical arrays of self-assembled QDs.

The first part of the paper describes a possible global architecture for QD-based quantum computers.
In the second part we will  analyze different possibilities for implementing information transfer along the QD-based bus, which  exploit the global dynamics of the system. We shall devolve particular attention to the constraints imposed on the scheme by the {\it physical} nature of the system itself.
Later we shall discuss the robustness of the proposed schemes  
and compare the performances of our quantum bus against the use of a microcavity.

\section{Quantum computer hardware design}\label{hard-des}
Self-assembled quantum dots are characterized by different properties, depending if we are considering an ensemble of {\it in-plane} QDs (the plane being orthogonal to the direction of growth)
or an array of QDs stacked in the growth direction.
In the following we will refer to the latter structure as 'Vertically Stacked Array' (VSA).

VSAs have very distinctive characteristics. First of all  they tend to form naturally due to strain propagation\cite{QD}. 
Secondly  self-assembled quantum dots are strongly confined in the growth direction, due to bandgap mismatch between the different semiconductors. This implies that (i) particle tunneling between nearby, vertically stacked QDs can be more difficult, even for small QD-QD distances\cite{PRBSergiolast} (ii) QDs can be very close together and
(iii) interactions between vertically coupled quantum dots may be quite strong (see e.g. Ref.\onlinecite{PRBBiolatti}). The latter characteristic is
very promising since quantum operations based on these  interactions can be carried out in times much shorter than the relevant decoherence times\cite{PRBBiolatti}.     
Last but not least, due to the  strain propagation, vertically stacked QDs tend to be noticeably different in size, which make them good candidates for energy selective addressing of specific excitations, a key ingredient in many QDs-based QIC proposals\cite{PRBBiolatti,PRBSergio,Paulibl}.

On the other side in-plane quantum dots can be produced in fairly uniform size  by controlling  the growth parameters: at present samples can be  already mono-disperse up to only few percent size variation.
This uniformity will be most probably  improved much further in the near future, especially since another important technological application, such as QD-based lasing, requires monodispersity\cite{lasing}. In addition there is already a good control of the in-plane QD density (which can be varied between $10^9$ to $10^{12}$ cm$^{-2}$) and, at the moment, techniques, e.g. 'seeding',  to produce more regular in-plane distributions are proposed and  tested\cite{seeding}.

\begin{figure}
\putfig{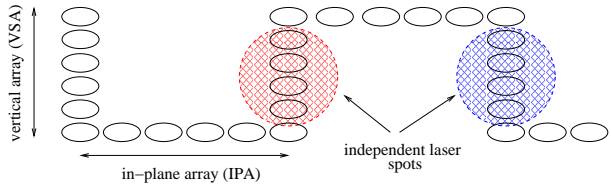}{8.0}
\caption{Schematic sequence of vertical arrays (VSA) and in-plane arrays (IPA)
of QDs. Different VSAs can be addressed by different laser spots.}
\label{fig1}
\end{figure}

The differences outlined above, between VSA and in-plane QDs, suggest us to diversify the use of  the two categories,  when thinking to the hardware structure of a quantum computer.
The structure we envisage in this paper is composed by a series of VSAs - for example laying in the same plane - connected by chains of in-plane QDs, which we will refer to as In-Plane Arrays (IPAs, see Fig.~\ref{fig1}).

This design allows for compact implementation of a quantum computer, since in principle it  may confine the hardware  in a single slab. This architecture 
satisfies the request of   easier lateral access (e.g. for wiring) and manipulation of each QD forming the computer (e.g. for applying external fields to specific dots). This design is also {\it scalable}, in the sense that the same basic structure (VSA plus related IPA) could be repeated at will, as well as the slab structure as a whole. 

We underline that the hardware structure we propose has yet  another very advantageous  feature. It can be built such that (i) each quantum register (i.e. any VSA) would be addressable by energy selective schemes (exploiting the natural diversity of stacked QDs), as required by many QIC schemes\cite{PRBBiolatti,PRBSergio,Paulibl}, but, at the same time, (ii) different quantum registers (different VSAs) could be also discriminated {\it spatially}, e.g by near-field techniques.
This can be achieved  
by making the quantum buses (i.e. the  IPAs),  longer or of the order of 1000\AA (see Fig.~\ref{fig1}). In this way the different VSA
can be spatially resolved by different laser pulses. 
This is a big improvement in respect to architectures which use  energy selective addressing only:
in our proposal even if the same laser color would in principle couple to one QD per VSA, we would be still able to address these resonant quantum dots {\it separately}. In general we could identify the QD sequence in each VSA with a {\it short}, well defined laser color sequence, and this sequence  could then  be repeated (totally or partially) for different VSA.
In principle the whole hardware structure could be built as a stacked sequence of mono-disperse QD planes, subsequently e.g. etched to form the VSA-IPA chosen structure.
This combined spatial-plus-energy selective addressability strongly enhances the hardware flexibility and helps overcoming
 the problems related to schemes which use energy-selective addressing only (e.g. NMR-based schemes).

\section{Quantum-dot quantum-bus potential implementations}
VSAs have been analyzed in details in many QIC proposals\cite{PRBBiolatti,PRBSergio,Paulibl}; in this paper
we will  concentrate on possible ways of implementing a quantum-bus by using an IPA as hardware.

We will discuss in details two potential implementations 
which   exploit  the {\it global dynamics} of excitons in the IPAs.
In both cases we shall (i) consider that state-of-the-art in-plane QDs can be made mono-disperse up to  few percents and (ii) discuss a way to fine-tune  QD-QD interactions  {\it a posteriori}.

\subsection{Implementation by global dynamics: general  considerations}\label{gen-phys}
Recently
there have been  interesting developments in the study of the global dynamics of a spin-chain\cite{Datta,Bose,gl-dyn,Montangero} as a possible way of transferring quantum information across the chain.
It is well known that the time necessary to transfer information along a spin or pseudo-spin chain in QIC schemes based on nearest-neighbor gating operations (e.g. see \Refon{PRBBiolatti,PRBSergio,Briggs,Paulibl})
scales with the number of spins.
There is though a widespread feeling that, when the transfer of information along a spin chain is based on the chain 
global dynamics, the transferring time can be independent from the chain length.

We will explicitly show that, for real systems,   the  time $t_t$ for transferring a qubit of information across a spin chain still depends on the chain length (for our specific hardware the number of QDs forming the IPA). 
In some cases though and for the same chain length, $t_t$
 may be smaller when exploiting global dynamics than by using nearest-neighbor gating operations, and some other advantages may occur in the first case.

To understand why this is so, we must take into consideration some very general constraints, which are  due to the fact
that any  hardware systems is indeed a physical system. These constraints are often overlooked in more 
abstract descriptions of spin-chain global dynamics;
when 'translating' the schemes into a physical system though, physical limitations must be taken into consideration.

Let us consider a spin-chain global dynamics based on nearest-neighbor interactions $J_{i,i+1}$'s which are {\it modulated} along the chain,
\be J_{i,i+1}=J\cdot f(N),\label{genJi}
\ee 
where $J$ is a constant representing a typical energy scale of the system and $f(N)$ is the function describing the coupling modulation along the chain. We assume  the modulation to be 
a reasonably regular function and to be un-bounded,  so that $f$ will depend over the chain length $N$.
 
The physical constraints in applying this scheme to any concrete  system  are connected to the following observations: 
(i) in any physical system there will be an upper limit $J_{max}$ to the interaction  strength, i.e. 
\be max\{J_{i,i+1}\}\le J_{max} \label{Jmax}\ee 
and (ii) for practical purposes the transfer time  $t_t$ must be smaller than the typical decoherence times, i.e. 
\be t_t< \tau_{dec}.\label{dec}\ee
 We shall show that Eqs.~(\ref{Jmax}) and (\ref{dec})  (i) imply that the transfer time $t_t$ across the chain depends over the chain length $N$ and that (ii) this sets a physical limitation to the maximum length of the chain we can consider.

From \Eq{Jmax} and \Eq{genJi} we have 
\be J\cdot f(N)\le J_{max}.\label{1stc}\ee
 Since $J/\hbar$ is the typical frequency of the system, we can assume that
$t_t\sim (J/\hbar)^{-1}$. From \Eq{1stc}
we obtain
\be {f(N)\over J_{max}}\le t_t, \label{gent}\ee
which shows that (i) the transferring time has a lower bound and, more important, (ii) this lower bound (and hence the transferring time) depends on the chain length $N$.
Using now \Eq{dec}, we obtain
\be {f(N)\over J_{max}}<\tau_{dec}, \ee
which sets a {\it physical} upper limit to the mathematically un-bounded $f(N)$ and hence to the chain length $N$.

For chains in which the modulation of the couplings $\{J_i\}$ is bounded, the dependence of $t_t$ on the chain 
length $N$ is due at least  to the constraint that information cannot travel faster than light. This is clearly a much weaker  constraint, and, as we shall see below, our calculations show that, even for the extreme case of chains with un-modulated couplings,
 much stringent relations between $t_t$ and $N$ usually occur. 

Among the various  proposals related to spin chain global dynamics, we will first consider the one
described in \Refon{Datta}. 
This is a scheme for perfect transfer of information (a quantum state or two entangled states) across a spin chain, by exploiting the global time evolution of the system.

The requirement for perfect transfer is that the system is governed by
 a XY Hamiltonian, in which the non-diagonal couplings between nearby spins $J_{i,i+1}$ must satisfy certain sequences, e.g.\cite{Datta} 
\be\label{Jis}
J_{i,i+1}=J\sqrt{i(N-i)},
\ee where $N$ is number of spins in the chain and $i=1,...N-1$. \Eq{Jis} represents an unbounded modulation of the coupling strengths of the type considered in \Eq{genJi},  which is in addition
symmetric in respect to the centre of the spin chain.
The time requested for the perfect transfer is\cite{Datta}
\be t_{(n+1)pk}=(2n+1){\hbar\pi\over 4J},\label{tnpk}
\ee $n=0,1,2...$, and it
 is determined by the value of $J$.

In particular 
let us consider the transfer time corresponding to $n=0$,  $t_{1pk}=\hbar\pi/(4J)$ and rewrite \Eq{Jis} as
 \be J=J_{max}/\sqrt{int(N/2)\cdot[N-int(N/2)]}, \label{J}\ee where we have optimistically assumed that at the center of the chain $J_{i,i+1}=J_{max}$. Notice that the latter equation implies that
\be t_{1pk}\sim {N\over 2}{\hbar\over J_{max}},\label{t1pk}
\ee i.e. there is direct proportionality between minimum time of transfer and chain length $N$. 
If we consider a transfer across a qubit chain done by nearest-neighbor two-qubit gates, and the ideal situation in which all the physical couplings used in the gates are given by $J_{max}$, we would have
\be t_{1pk}\sim N{\hbar\over J_{max}}.
\ee  
Comparing the last two equations, we see that the actual gain in using the global dynamics of a spin chain, when the {\it physical condition} of including a maximum strength for a coupling is taken into account, is a factor 2, coming from the symmetry of the modulation, but the transfer time still
scales with the total number of spins N. Fig.~\ref{perf_tranf}, lower panel, shows  
$t_{1pk}$ extracted from the simulation of spin chains with nearest-neighbor couplings following \Eq{Jis} in respect to $N$. 
It confirms the direct proportionality between minimum transfer time and chain length.

The concrete advantage of using global dynamics over two-qubit gating becomes then the possibility of simplifying the implementation, an issue that should not be underestimated, but may again depend over  the specific physical characteristics of the system chosen as hardware. At the very least, exploiting global dynamics lifts the necessity of driving every single two qubit operation and this consequently avoids potential sources of errors; but
what is really important when considering a spin chain global dynamics as a way to implement a quantum bus, is to maximize  the 
simplicity of the device.

The fact that, as shown above,  for physical realizations,   $N$ cannot be arbitrarily long, implies that
the properties  relevant to many practical implementations will be  the ones of medium or relatively short chains, while previous literature  has mainly focused on the large-N limit\cite{Bose,Montangero}.
 
With this in mind, in the following we shall focus on chains characterized by $N\stackrel{<}{\sim}10$, both underlining some specific general properties 
 and referring to a specific physical system -- the IPA-chain formed by QDs with excitons as pseudo-spins.

\subsection{F\"orster effect}
We will consider implementations of quantum buses in which 
the coupling between nearby in-plane QDs is provided by the F\"orster effect\cite{Foerster1,Foerster2}. Due to this coupling,
the excitation corresponding to a correlated electron-hole pair (exciton) can be transferred  from one QD to its neighbor. 

This effect has been observed as a powerful motor for excitonic transfer in colloidal self-assembled quantum dots\cite{Foerster2}. The measured strength of the F\"orster coupling in CdSe QDs is of about 0.3 meV, but it is believed that in structurally optimized systems the interdot energy transfer can approach picosecond time scales, i.e. meV coupling strength\cite{Foerster2}. 
Being  the structures of interest to this paper (self-assembled QD) of similar size in respect to colloidal QD and being typical values for QD-QD interactions of few meV\cite{PRBBiolatti}, 
 we will assume in our calculations 1 meV as a  reasonable {\it upper limit} for F\"orster interaction in  self-assembled QDs.

In the envelop function approximation, the F\"orster coupling between an exciton in QD$_i$ and an exciton in QD$_j$, $V_{Fij}$,  is given by\cite{Lovett2003} 
\ba
V_{Fij}&=&{O_iO_{j}e^2\over \epsilon R_{ij}^3}[<{\bf r}_i><{\bf r}_{j}> \nonumber \\
&-&{3\over R_{ij}^2} (<{\bf r}_i>\cdot {\bf R}_{ij})(<{\bf r}_{j}>\cdot {\bf R}_{ij})] \label{V_F}
\ea 
where
\be
O_i=\int\psi_e^i({\bf r})\psi_h^i({\bf r})d^3r
\ee
is the overlap between the hole and electron envelop functions in  QD$_i$, $R_{i,j}$ is the distance between QD$_i$ and QD$_j$, and $e<{\bf r}_i>$ is the dipole moment due to the atomic part of the wave function in QD$_i$.

We underline that no particle tunneling is involved in the F\"orster process, but only energy transfer:
 it depends in fact on $O_i$, the overlap of  
electron and hole wave-functions  in the {\it same} QD. In tunneling,
 the relevant quantity is instead the overlap of wave-functions belonging to different QDs.
Due to energy conservation, this coupling is most effective between 
very similar QDs.
When QD chains are involved though,  we can picture a dynamical F\"orster-based  process between far-away but similar QDs positioned along  the chain. The similarity of the first and last QD in the chain ensures that the total process of transferring the excitation across the IPA remains energy preserving. 
 
Labeling with $|0\rangle_i$ ($|1\rangle_i$) the absence (the presence) of a single ground-state exciton in QD$_i$,  
the Hamiltonian for a chain of N QDs, in which we consider at most single exciton occupancy of the IPA, is given by
\ba H &=& \sum_i^N E_i |1\rangle_i\langle 1|_i \nonumber \\ &+&
 \sum_{i=1}^{N-1} \left( V_{Fi,i+1}|1\rangle_i\langle 0|_i\otimes |0\rangle_{i+1}\langle 1|_{i+1}+h.c.\right),\label{HF}\ea
where we have assumed no single particle tunneling.  Contributions behind nearest-neighbors are second order,
since the F\"orster effect scale as $R_{ij}^3$,  
and have been neglected.

\subsection{Implementation by global dynamics using a QD chain: first method (modulated coupling)} 
For concreteness, in the following we shall consider as chain an IPA, and as  (pseudo) spins degrees of freedom  the presence or absence of a  ground state exciton in each QD\cite{PRBBiolatti}.
The transfer of quantum information across the  ``spin chain'', will then correspond to the transfer of  an excitonic state across the IPA, by means of exploiting the global excitonic dynamics of the system. 
 The general conclusions on the IPA can though be easily transferred to other types of physical hardware.

The method we are going to discuss in details represents one of the possible
 schemes for  transferring information between separate VSAs in the computer architecture previously outlined.
To be able to address different VSAs with independent laser pulses, 
we  require the length of the IPA to be longer than $1000\AA$. A reasonable distance between the center of nearby QDs is at least of the order of $150-200\AA$. This implies that the IPA must be formed by at least 5-7 QDs. The question is now if excitonic transfer through such a chain can be performed in a reasonable time  and with reasonable accuracy.

In our system the (pseudo) spin degrees of freedom are represented by the presence ($|1\rangle_i$) or absence ($|0\rangle_i$) of a ground state exciton in each QD$_i$, where $i=1,N$ labels 
sequentially the QDs in the chain. This is the same  encoding proposed in Ref.~\onlinecite{PRBBiolatti}. The Hamiltonian of the system is given by \Eq{HF}.

\begin{figure}
\putfig{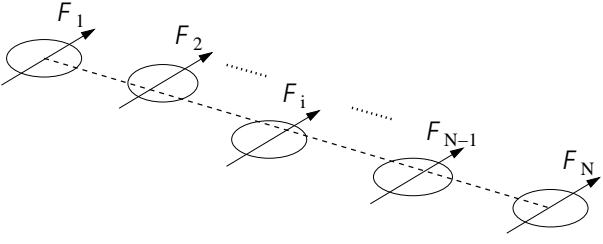}{8.0}
\caption{Sketch of an IPA in which an external static electric field $\cal{F}_i$ has been applied to every QD$_i$.
}
\label{IPAandFields}
\end{figure}

We shall refer to the scheme presented in \Refon{Datta} as the Modulated Coupling (MC) scheme. 
In respect to the XY Hamiltonian required by \Refon{Datta},  \Eq{HF} presents diagonal, in general not-equal, terms $E_i$ and XY non-diagonal terms, provided by the  F\"orster coupling $V_{Fij}$, which in general do not satisfy \Eq{Jis}.
To obtain a good approximation to the 'mathematical' Hamiltonian, we need  at least to
engineer  the non-diagonal terms into the required sequence Eq.~(\ref{Jis}).
From Eq.~(\ref{V_F}),  we see that we can tune the value of the  F\"orster coupling by (i) engineering the value of the interdot distance $R_{ij}$ or (ii) by tuning the value of the overlap integrals $O_i$.
Since roughly $V_{Fij}\sim 1/R_{ij}^3$, 
 to  engineer $V_{Fij}$ according to Eq.~(\ref{Jis}) over an IPA of 5-7 QDs  with $R_i\sim 150-200\AA$,
implies to be able to experimentally control $R_{ij}$ up to few Amstrongs, which is at the moment experimentally very challenging.
A more flexible approach, is to vary, {\it a posteriori}, the electron-hole function overlap $O_i$. This is doable by applying a suitable (but static) {\it in-plane} electric field to each QD, as described in \Refon{PRBBiolatti}. The field is to be applied in the in-plane direction orthogonal to the IPA's direction (see \Fig{IPAandFields})
.
The advantage of this scheme is to allow also for corrections of  possible errors occurred in the growth process, since the applied electric field can be varied {\it a posteriori}, after the growth process is completed.

Let us consider a 'uniform' initial condition, in which $V_{Fi,i+1}= V_{Fj,j+1}$ for every $i,~j$.
If $N= 5-7$  and $\{V_{Fi,i+1}\}$ has to satisfy \Eq{Jis}, the maximum correction required over the coupling strength $V_{Fij}$ is of
 about 30\%. This corresponds to a 20\% variation over the overlap integrals $O_i$'s. We need now to estimate the required electric field, in order to asses the feasibility of the scheme.
If we assume Gaussian electron and hole wave-functions, the relation between $O_i$  and the distance $d$ between the wave-function centers is
\be
O_i(d_i)=2^{3\over 2}\Pi_{t=x,y,z}\left[\sqrt{\lambda_{te}\lambda_{th}\over  \lambda_{te}^2+\lambda_{th}^2}\right]_i\exp\left(-{1\over 2}{d_i^2\over\lambda_{xe,i}^2+\lambda_{xh,i}^2}\right),\label{O_i}
\ee 
where $\lambda_{te(h)}$ are the characteristic length of the wave-functions in the $t=x,y,z$ directions.
If we assume similar dots (a reasonable request for an IPA), we obtain 
\be
d_i\approx \sqrt{2(\lambda_{xe}^2+\lambda_{xh}^2)\left|\ln\left({O_i\over O_{np}}\right)\right|},\label{d_iO}
\ee
where $O_{np}$, is the unperturbed value of the overlap integral, i.e. the value it assumes for the central(s) QD(s) which is coupled to the neighbors by  the coupling constant with the the highest value  (see \Eq{Jis}).
On the other side $d_i$ can be also related to the static electric field ${\cal F}_i$ applied to QD$_i$ by\cite{PRBBiolatti}  
\be d_i=e{\cal F}_i[(1/m_e\omega_{e,i}^2)+(1/m_h\omega_{h,i}^2)].\label{d_iE}\ee
Assuming typical GaAs parameters ($m_h=0.34m_0$, $m_e=0.067m_0$, $\hbar\omega_h\sim20$ meV, $\hbar\omega_e\sim40$ meV), and using Eqs.~(\ref{d_iO}) and (\ref{d_iE}), we see that a variation of 20\% over the overlap integral $O_i$ 
 can be produced by a field of the order of 30KeV/cm, perfectly reasonable from the experimental point of view. We can then assert that the scheme we are going to describe more in details could in principle be implemented by state of the art technology. 

\subsection{Modulated coupling: simulations and errors}

In this section we shall discuss and simulate the excitonic dynamics  in  IPAs and discuss the effect of possible sources of errors for the transmission of the excitonic state across the IPA.
Because of the generality of most of the results, we will indicate the couplings as $J_{i,i+1}$, keeping in mind that when the considered hardware is an IPA the coupling is given by the F\"orster effect \Eq{V_F}.
We will consider in our calculations realistic  QDs parameters and in particular GaAs-based IPAs with $N\le9$ and $J_{max}\le 2$meV, $J_{max}$ the {\it upper limit} for  the coupling between nearby QDs. 

\begin{figure}
\putfig{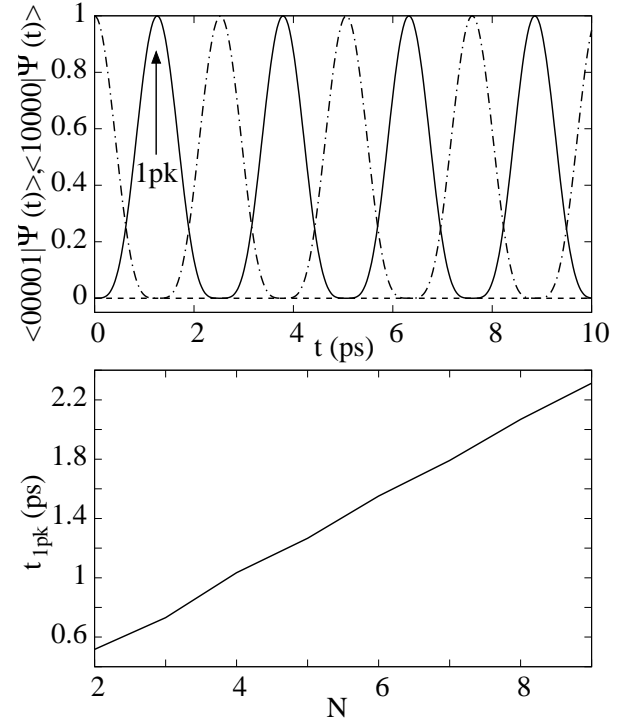}{8.0}
\caption{Upper panel: $\langle10000|\Psi(t)\rangle$ (dashed-dot line) and  $\langle00001|\Psi(t)\rangle$ (solid line) vs time for N=5. '1pk' marks the first resonance position.
Lower panel: $t_{1pk}$ vs N.  
}
\label{perf_tranf}
\end{figure}
Let us first of all consider  the case in which the conditions in \Refon{Datta}  are perfectly satisfied. 
Fig.~\ref{perf_tranf} (upper panel) shows a simulation of the dynamics of the system for $J_{max}=1$meV and an IPA of 5 QDs. The state at time $t=0$
 is 
$|\psi(0)\rangle=|10000\rangle$, i.e. single exciton in  QD$_1$. The projection of the  state of the
system $|\psi(t)\rangle$
at time $t$ on the initial and the desired final state (i.e. exciton in QD$_N$, the last QD), $\langle10...0|\psi(t)\rangle$ and
$\langle0...01|\psi(t)\rangle$ respectively, are plotted.
As predicted in \Refon{Datta}, there is periodic perfect transfer between the two states, the first transfer occurring at $t_{1pk}\approx 1.3 $ps.
The lower panel presents the plot of the numerical value obtained for $t_{1pk}$ when the length of the IPA is changed.
As can be seen, the linear dependence of \Eq{t1pk} is recovered.
Notice that, even for $N=9$, the transferring time remains very short, $t_{1pk}\approx 2.3 $ps, confirming that, even if there is direct proportionality between $N$ and  $t_{1pk}$, at least for relatively short chains, the effect of this proportionality is not dramatic.

Let us compare these results with the use of microcavities, which is
the alternative route for transferring information at 'long distance' in QD-based quantum computers.
 If we consider a {\it comparable} transfer done by mean of   a microcavity, i.e. with a coupling of 1 meV between QDs and cavity, the minimum time necessary for the transfer  is greater than 100 ps\cite{Mang1}.
Extrapolating from  \Fig{perf_tranf}, lower panel, and assuming the relation \Eq{t1pk}, a transferring time $t_{1pk}=100$ps  would correspond to a transfer over an IPA with $N\approx 400$, i.e., assuming $R_{ij}=150\div 200\AA$, over
a distance of $6\div 8\mu m$.  We see than that, from the transfer-time point of view,
 the use of microcavities becomes advantageous only when really long distances, of the order of several 
microns,
 are involved. For  these distances  the design and modulation of long spin-chains (long IPAs) may also 
become too complicated.  

We notice also that a phase shift is originated at the end of the transfer, depending on the number of QDs N\cite{Datta}. In particular when $N-1$ is an exact multiple of 4, no phase shift is produced\cite{Datta}.
Since, once the length of the chain is known, this fixed shift can be corrected by a simple single qubit
phase rotation operation done after  the transfer, we will neglect the effects of this shift in the following discussion.   

\begin{figure}
\putfig{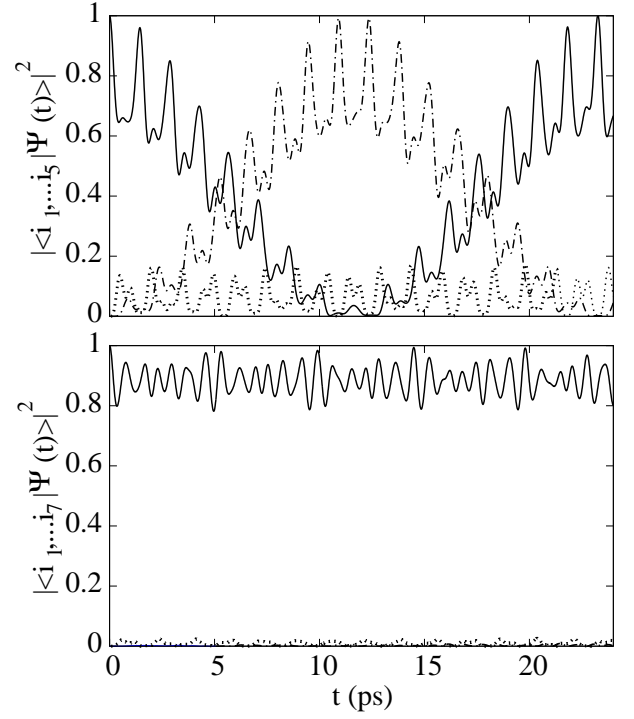}{8.0}
\caption{Upper panel: $|\langle10000|\psi(t)\rangle|^2$ (solid line), $|\langle00001|\psi(t)\rangle|^2$ (dashed-dot line) and $|\langle00100|\psi(t)\rangle|^2$ (dotted line) vs time, for $J_{max} = 1$ meV, $\hbar\omega_{ex,i}=20$ meV and  $\omega_{ex,i}/ \omega_{hx,i}= 3$ and GaAs parameters. \\ Lower panel: $|\langle1000000|\psi(t)\rangle|^2$ (solid line), $|\langle0000001|\psi(t)\rangle|^2$ (dashed-dot line) and $|\langle0001000|\psi(t)\rangle|^2$ (dotted line) vs time. Same parameters as upper panel.}
\label{starktransf}
\end{figure}

Let us focus now on our specific system, the IPA.
If we consider as a starting point a chain of similar QDs regularly spaced, and approximate the electron and hole wave-functions as Gaussian, we can estimate the Stark shift of the excitonic level of QD$_i$ in respect to the central,
 unperturbed, QD$_{np}$ produced by the application of the fields $\{\cal{F}_i\}$ to satisfy \Eq{Jis} (see \Fig{IPAandFields}). This is given by
\be \left|E_i\right|=-\hbar\omega_{ex,i}{1+{m_e\omega_{ex,i}\over m_h\omega_{hx,i}} \over
 1+{m_e\omega_{ex,i}^2\over m_h\omega_{hx,i}^2}}\ln{J_i\over J_{max}},\label{E_Stark}
\ee
where $J_{max}=max\{J_i\}$ is the value of the coupling between QD$_{np}$ and its neighbors. The ratio $J_i/ J_{max}$ can be calculated  from Eq.~(\ref{Jis}). 
For a given semiconductor, \Eq{E_Stark} depends on the ratio  $\omega_{ex,i}/ \omega_{hx,i}$, but for the relatively short IPAs we are considering, we see that, even when  $\omega_{ex,i}/ \omega_{hx,i}\approx 3$ and $\hbar\omega_{ex,i}=20\div 40$ meV,   $|E_i|>1$meV. 
When we consider the approximate sequence $\{E_i\}$ as given by Eq.~(\ref{E_Stark}) and 
let the system evolve according to the Hamiltonian in \Eq{HF}, our simulations show
 that the excitonic transfer from the first to the last QD may be achieved, but depending on the  parameters chosen. In \Fig{starktransf} we compare the case of N=5 (upper panel) and N=7 for the same structural parameters (see figure caption)
and $|\psi(0)\rangle=|10...0\rangle$. Even when the transfer is achieved (N=5), the periodic pattern becomes more complex and other excitonic states of the system are populated during the transfer itself (e.g. the probability of occupying state $|00100\rangle$ is shown). For N=7, over the same interval of time, no relevant transfer across the chain is achieved, while the exciton remains mainly localized in QD$_1$. 

We can think of improving this result for relatively short IPA by requiring careful tuning of at least part of the distances $R_{ij}$ between QDs and/or of other QDs parameters during the growth process, in order to obtain as a starting point 
 a sequence of $J_i$ which matches more closely the requirement of Eq.~(\ref{Jis}).  This implies the application of {\it smaller static electric fields} and consequently smaller Stark shifts. The energy shifts will now be in general randomly distributed, depending on how well the single coupling $J_{i,i+1}$ satisfies Eq.~(\ref{Jis}).

We have tested the robustness of this scenario by 
calculating the fidelity of the transfer across
a QD chain in which the sequence $\{E_i\}$ is produced randomly, with the constraint $E_i\le\delta E$.
The fidelity of the transfer for each set $\{E_i\}$ is calculated according to Ref.~\onlinecite{Bose} as
\be
\langle F\rangle_{Bl}={|f_N|\cos\gamma\over 3} +{|f_N|^2\over 6}+{1\over 2},\label{FBl}
\ee
with $f_N=\langle 0....1|\exp(-iHt)|1....0\rangle$, $H$ given by \Eq{HF} and  $\gamma=arg\{f_N\}$. In the following we will
set $\cos\gamma=1$ since we can always correct the phase shift after the transfer has occurred, as mentioned before. Notice that the
 fidelity in \Eq{FBl} includes the  averaging of the initial state of QD$_1$ over the whole Bloch sphere\cite{Bose}.
The obtained result will be then averaged (when appropriate) over different random configurations $\{E_i\}$ ($\langle F\rangle_{Bl,cf}$).
In this respect, when $\delta E \stackrel{>}{\sim} J_{max}$ it is important to perform this average 
over some thousand configurations, in order to obtain reliable results.

\begin{figure}
\putfig{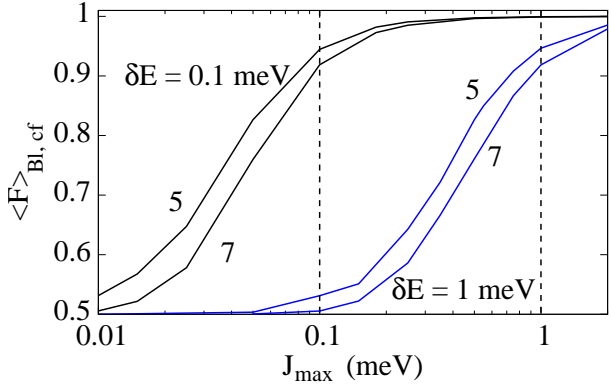}{8.0}
\caption{Average fidelity  $\langle F\rangle_{Bl,cf}$ vs  $J_{max}$, for $\delta E = 0.1$ meV and $\delta E = 1$ meV and chain length of 5 and 7 QDs, as labeled. The vertical dashed lines indicates the values for which $J_{max}=\delta E$.
}
\label{runIV_V}
\end{figure}

Results for IPAs with N=5 and N=7 are plotted in  Fig.~\ref{runIV_V}.
$\langle F\rangle_{Bl,cf}$ is plotted  in respect to the value of $J_{max}$, for two different values of $\delta E$.
The behavior for the two values is similar, the relevant parameter being the ratio 
$\delta E/ J_{max}$
In particular, when $\delta E\approx J_{max}$, the fidelity is still greater than 90\%, but it 
 decreases  drastically for increasing   $\delta E/J_{max}$. The variation with the chain length is instead not substantial, though shorter chains perform better.
Notice that in \Refon{Montangero} the {\it opposite} trend is predicted, i.e. longer chains more robust than shorter ones.
 This is due to the fact  that in their simulations they use different constraints:
they vary $N$ {\it for fixed} time transfer $t_{1pk}$, i.e. for fixed $J$. 
In other words the maximum value  for the coupling between nearby spins is not fixed, and $J_{max}$ is let increasing with the number of spins $N$. 

When $\delta E\stackrel{>}{\sim} J_{max}$, the fidelity of the transfer $\langle F\rangle_{Bl}$  from first to last QD largely varies according to the specific set $\{E_i\}$. In addition
the presence of $\{E_i\ne 0\}$ modifies  the time at which $\langle F\rangle_{Bl}$
is maximum  (the same  occurs for the second method we shall consider, see inset of \Fig{runVII_VIII}).
In this respect a possible way to improve the fidelity would be  to initially measure the optimal transfer time for the specific set $\{E_i\}$
corresponding to the sample and then tune the  time for the transfer operation accordingly. For IPAs the main contribution to $\{E_i\}$ is given by growth and/or applied static electric fields,  so that measuring the optimal transfer time for the specific hardware piece might recover a significant part of the fidelity.
Nowadays tests on single electronic components are done routinely to check their specifications; we foresee that analogous  testing procedures will be needed for the various hardware components of a quantum computer.

\subsection{Implementation by global dynamics using a  QD chain: second method (flat coupling)}
We have stressed in Sec.~\ref{gen-phys} that the main advantage of using global dynamics should be to simplify the implementation of the computation.
For  implementing the MC method   with IPAs, carefull tuning of the QD properties is required both in the growth phase and a posteriori,
with the risk that large values of $E_i$ may result, affecting the fidelity. 
In order to simplify the implementation and overcome some possible
 sources of error, we will consider in this section the global dynamics of a spin chain, when all couplings between nearby spins are the same, i.e. $J_k=J_i$, and $E_k=E_i$ for every $i,k$.
We shall refer to this as ``flat coupling'' (FC) scheme.
 As underlined in Sec.~\ref{hard-des} these requirements should be easier to match, at least
 considering implementations which use IPAs as hardware, since mastering these  QD properties is already on the agenda of many experimental groups\cite{lasing,seeding}.

We want to address the questions (i) if transfer of a quantum state across the chain by exploiting global dynamics 
 is still possible and with which degree of accuracy  and (ii) how the results are affected by imperfections.

We consider again relatively short spin chains,  with $N=1,...9$, and simulate the evolution of the system according to the Hamiltonian \Eq{HF}.
A similar problem have been analyzed in \Refon{Bose}, but under different constraints and mainly considering the limit of long spin chains, so that properties of the relatively short chains which are relevant to the present proposal  have not been discussed. 

\begin{figure}
\putfig{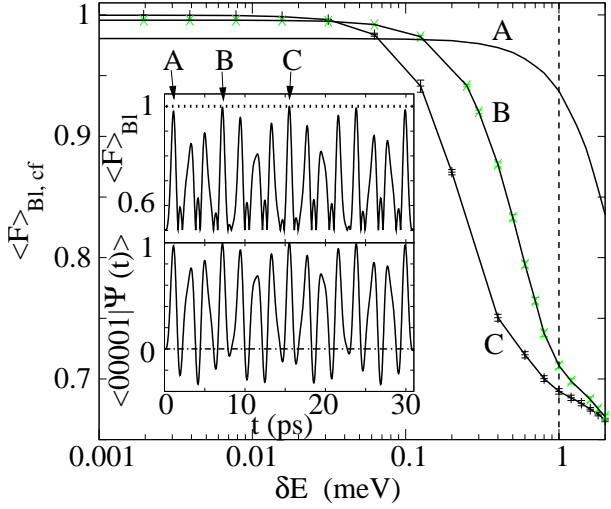}{8.0}
\caption{$\langle F\rangle_{Bl,cf}$ vs $\delta E$ for  three different resonances (labeled A,B and C), for N=5 and $J_{max}=1$ meV. The dashed vertical line corresponds to the value of  $J_{max}$. Inset: upper panel: $\langle F\rangle_{Bl}$ as a function of time for $\delta E=0$, 
N=5 and $J_{max}=1$ meV. Lower panel: real (solid line) and imaginary (dashed-dot line) part
 of $\langle 00001|\Psi(t)\rangle$ for $|\Psi(0)\rangle=|10000\rangle$ and same parameters as upper panel.
}
\label{runVII}
\end{figure}

\begin{figure}
\putfig{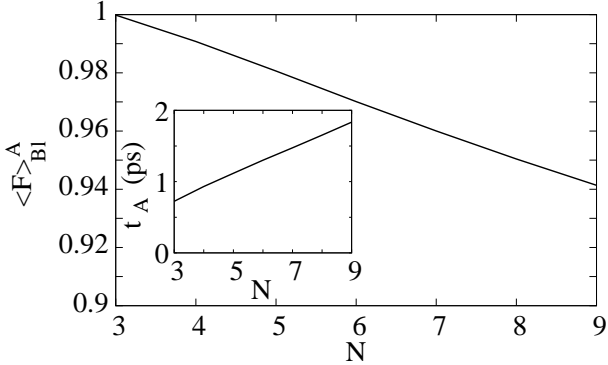}{8.0}
\caption{$\langle F\rangle^A_{Bl}$ vs N.  Inset:
$t_A$  vs N. Parameters as inset of \Fig{runVII}.
}
\label{runVI}
\end{figure}

The inset of \Fig{runVII} shows the time evolution of a N=5  chain with $E_k=0$, $k=1,N$ and $J_{max}=1$meV. The 
upper panel presents the fidelity   $\langle F\rangle_{Bl}$ in respect to time; the lower panel shows the real (solid line) and the imaginary (dashed-dot line) components  of  the projection $\langle 00001|\Psi(t)\rangle$ in respect to time, with
 initial conditions $|\Psi(0)\rangle=|10000\rangle$.
As can be seen, the time evolution {\it as a whole} presents a periodic behavior (within the numerical error), centered on the peak labeled with 'C'. The period  is proportional to  $1/J_{max}$. Within this period many {\it high-fidelity resonances} are present, some of which correspond to a fidelity $\langle F\rangle_{Bl}\sim 1$ (e.g.$\langle F\rangle_{Bl}^A=0.981$,  $\langle F\rangle_{Bl}^B=0.996$ and $\langle F\rangle_{Bl}^C=1.00$). 
The behavior just discussed represents
 a typical pattern for all the spin-chains here analyzed. In particular, for $N\le 9$, the very first resonance  (labeled 'A' in the inset of \Fig{runVII}) corresponds  always to a  fidelity $\langle F\rangle_{Bl}>94\%$ (\Fig{runVI}, main panel). 
In general chains with an odd number of QDs N have an overall period shorter than the corresponding (N-1) QD chains. Interestingly the phase of the projection $\langle 00001|\Psi(t)\rangle$ follows for FC the same dependence on N as for MC, i.e. it is given by $(-i)^{N-1}$.

In the inset of \Fig{runVI}, the time $t_A$ at which the resonance 'A' occurs, is shown in respect to N. 
Both $t_A$
 and the corresponding fidelity $\langle F\rangle_{Bl}^A$
present a {\it weak} linear dependence with the length of the
chain N, with $t_A$ though still below 2 ps for N=9 (inset of \Fig{runVI}). We
stress that for the applications we are interested in, we need
a finite, relatively small number of QDs, so these results are
very encouraging.

In all cases analyzed, at least one of the  resonances following peak 'A'  corresponds  to an even higher  fidelity (e.g. B and C in \Fig{runVII}, for N=5). The possibility of accessing these stronger resonances as a mean of a better information transfer along the chain, will depend on the values of $J_{max}$, and of the typical decoherence times of the system, since, as already mentioned,
 the times corresponding to the different resonances  scale as $1/J_{max}$.

We shall now analyze how the different resonances are influenced by imperfections. As for MC, we have considered sets $\{E_i\}$
with $E_i$'s randomly distributed and $E_i\le\delta E$\cite{note2}. 
An average over an appropriate number of such configurations is taken to calculate the fidelity $\langle F\rangle_{Bl,cf}$.
We present the results in \Fig{runVII}, where the average fidelities $\langle F\rangle_{Bl,cf}^{K}$, $K=A,~B,~ C$, calculated at 
the  times $t_A$, $t_B$ and $t_C$ at which the resonances occur, respectively, are  plotted against $\delta E$.
Our results  show that the resonances B and C are less 'robust' in respect to the first resonance (A), i.e. they  tolerate a lesser amount of imperfections in the flat distribution of the energies $E_k$. 
In fact for very small $\delta E/J_{max}$,  $\langle F\rangle_{Bl,cf}^{B,C}\approx 1$, while $\langle F\rangle_{Bl,cf}^A=0.98$.
$\langle F\rangle_{Bl,cf}^{B,C}$ though decrease rapidly already for $\delta E/J_{max}\sim 0.1$, while  $\langle F\rangle_{Bl,cf}^A$ remains basically unperturbed.  For  $\delta E/J_{max}=1$, $\langle F\rangle_{Bl,cf}^{B,C}$ drop to about 0.7, while
$\langle F\rangle_{Bl,cf}^A\sim 0.94$, still a very significant  fidelity.

\subsection{Comparison between modulated and flat coupling}
\begin{figure}
\putfig{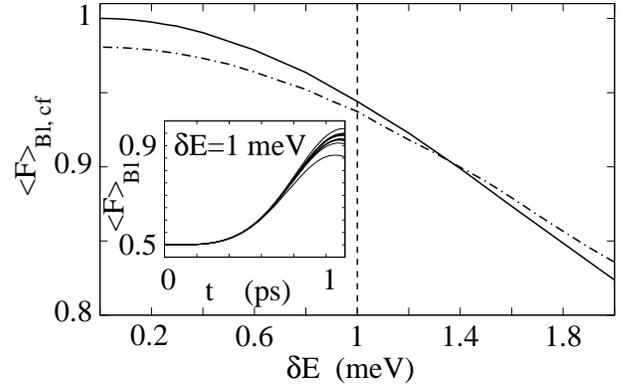}{8.0}
\caption{ $\langle F\rangle_{Bl,cf}$ vs $\delta E$ for MC scheme at $t_{1pk}$ (solid line) and FC scheme at $t_A$ (dashed-dot line). N=5, $J_{max}=1$ meV. Inset: $\langle F\rangle_{Bl}$ vs time for FC scheme and same parameter as main panel. The x-axis upper limit corresponds to $t_A$. 
}
\label{runVII_VIII}
\end{figure}

We shall now compare  MC and FC schemes.
In \Fig{runVII_VIII} we compare $\langle F\rangle_{Bl,cf}^{A}$  (dashed-dot line), with 
the average fidelity $\langle F\rangle_{Bl,cf}$ calculated at $t_{1pk}$ (see \Eq{tnpk})  with $J_i$'s given by \Eq{Jis}  (solid line). They are plotted  in respect to $\delta E$,
which is varied up to 2 meV, for N=5 and $J_{max}=1$ meV.
We notice that for  $\delta E/J_{max}\approx 1.3$, the fidelity of MC method becomes smaller, confirming
 that, if $\delta E\stackrel{>}{\sim}J_{max}$ is to be expected (e.g.  when using IPAs as hardware), it is convenient to opt for the simpler FC scheme.
The inset of  \Fig{runVII_VIII} presents $\langle F\rangle_{Bl}(t)$ for FC in respect to time, for $t\le t_A$ and  $\delta E=1$ meV. Each curve corresponds to a  
different  random set $\{E_i\}$. The plot shows that $\langle F\rangle^A_{Bl}$ is spread over a large interval which increases with increasing $\delta E$ (not shown).
Two things should be noticed, the first is that the fidelity corresponding to a specific $\{E_k\}$ configuration can be (clearly) greater than the average fidelity $\langle F\rangle_{Bl,cf}$, the second that the maximum  the fidelity for the transfer may occur at times  different from $t_A$. Similar conclusions are reached when analyzing  $\langle F\rangle_{Bl}(t)$ in the MC scheme.
These considerations imply that (i) operatively it will be important to carefully measure the sample characteristics in order
to optimize the transfer time, and/or, as possible alternative, (ii)  specific sets $\{E_i\}$  could be designed to optimize the fidelity.

FC scheme can be seen as representing a very bad implementation of the MC scheme,
in which  every trace of the sequence  \Eq{Jis} has been washed out.  
In this respect the occurrence of the very strong first resonance 'A'  -- resonance
which coincides, time and strength, with the first transfer  of the MC scheme
when $N\le 3$ -- suggests that,  for relatively  short chains, the MC scheme is very robust in respect to imperfections in the $J_i$'s sequence \Eq{Jis}. This robustness decreases with N, as the strength of $\langle F\rangle^A_{Bl}$ does (see \Fig{runVI})\cite{note}.
This interpretation also helps to understand  why (i)the same phase dependence on N is observed in the FC and in the MC scheme for the projection  $\langle 00001|\Psi(t)\rangle$ and (ii)$t_A$ has a linear dependence on $N$ similar to $t_{1pk}$ (compare \Fig{perf_tranf}, lower panel,
and inset of \Fig{runVI}).

Our results suggest,
 for simplicity of implementation and   especially if it is  possible to have access to the  resonances corresponding to 
the highest fidelities (B and C in \Fig{runVII})  the FC as the best scheme, at least when using IPAs as hardware.

\section{Summary and Conclusions}
We have proposed a potential implementation for a quantum computer  where both the 
computation  and  the communication between far away qubits utilize exclusively  quantum dot arrays. We have discussed how the different  
characteristics of in-plane and stacked arrays of QDs can be exploited to this purpose, the first as quantum buses and the second as quantum registers. The structure we propose is scalable and avoids the use of microcavities.

In the second part of the paper we have focused on the properties of the global dynamics of 'relatively short' spin chains, such chains been the most relevant in respect to the physical problem at hand.
By mapping ground state excitons into (pseudo) spins, this dynamics can be used as a way to implement long distance communication between far away QDs connected by an  in-plane array of QDs.
We have discussed and compared
two different potential schemes, the first based on the coupling between nearest-neighbor spins modulated along the chain\cite{Datta}, the second based on a flat, uniform coupling. 
In particular, in the case of flat coupling, 
we have shown that the system presents a behavior overall periodic in time,
containing a series of resonances 
  corresponding to a fidelity of the state transfer across the chain very close to 100\%.
The very first of these resonances corresponds always to a high fidelity, {\it larger than 94\%} for $N\le 9$. This fidelity presents a linear dependence on the spin chain length N, but such a dependence is relatively weak. 
According to the system physical characteristics (maximum value of the nearest-neighbor coupling and decoherence times) one or all of these resonances will be accessible experimentally and exploitable as a mechanism for transferring information. 
The effects of imperfections on these resonances has also been analyzed, showing that the very first one is very robust.
  
It is clearly possible to design many  more schemes to implement  information transfer along  in-plane QD chains -- schemes either based on the  global dynamics of the chain
 or on   nearest-neighbor two-qubit gating. A systematic analysis of  all possible schemes goes beyond the scope (and the possibilities) of this paper.
We hope that the two potential implementations here discussed represent a good starting point for a stimulating discussion on the all-QD-based hardware structure we have proposed.
\section*{Acknowledgments} We are grateful to M. Feng, N. Datta, M. Christandl, A. Bychkov, T.P. Spiller, and B.W. Lowett  for stimulating and fruitful discussions.


\begin{thebibliography}{99}
\bibitem{note2}By applying external electric fields, imperfections on the coupling constants may be corrected and partially transformed into imperfections on the $\{E_i\}$ set (as seen in the first part of the previous section), so for our purposes it is sufficient to study the effects of imperfections over  the energy set.
\bibitem{note}This conclusion is different from the one in \Refon{Montangero} for the same reasons as  for the effect of  imperfections over the set $\{E_i\}$,
\end{thebibliography}
\end{document}